\documentclass[10   xpt,conference]{IEEEtran}
\IEEEoverridecommandlockouts
\usepackage{cite}
\usepackage{amsmath,amssymb,amsfonts}
\usepackage{algorithmic}
\usepackage{graphicx}
\usepackage{comment}
\usepackage{textcomp}
\usepackage{xcolor}
\usepackage{url}
\def\BibTeX{{\rm B\kern-.05em{\sc i\kern-.025em b}\kern-.08em
    T\kern-.1667em\lower.7ex\hbox{E}\kern-.125emX}}
    
\begin{document}

\title{P4Kube: In-Network Load Balancer for Kubernetes}



\author{
 Garegin Grigoryan\\ grigoryan@alfred.edu  \\Alfred University 
 \and
Kevin Penkowski \\ kwp5892@rit.edu \\Rochester Institute of Technology 
  \and
Minseok Kwon \\ jmk@cs.rit.edu \\Rochester Institute of Technology 
}

\maketitle
\begin{abstract}
Kubernetes Services such as LoadBalancer and NodePort expose applications running on pods within a Kubernetes cluster to external users. While the LoadBalancer Service requires an external load-balancing middleware, its alternative, NodePort Service, adds additional hops on the path between clients and the worker nodes. In this paper, we propose P4Kube, a framework consisting of a P4 data plane program and a Kubernetes plugin. Our solution effectively performs load balancing of requests to the worker nodes of a cluster based on the number of running replicas. In P4Kube, the data packets completely bypass the system's control plane. Unlike the previous work, to update its state, the P4Kube data plane works directly with the Kubernetes control plane without any involvement of the network control plane. Our experiments show up to 50\% improvement in the average request time to the cluster compared to conventional approaches.
\end{abstract}

\section{Introduction}
\label{sec:intro}
Kubernetes~\cite{kube} is a dominant tool for container orchestration~\cite{orchestrators}.  %Containerization is a method of bundling an application and its dependencies in a standardized way to allow for easy and consistent deployment and running of an application. Container orchestration tools provide a framework for the deployment, management, scaling, and monitoring of clusters of containerized applications.
A Kubernetes Control Plane server (CP) organizes one or more containerized applications into different pods, the smallest unit deployable on the worker nodes. A Kubernetes Service~\cite{services} allows access to Kubernetes pods via exposed TCP ports. %, a must be deployed via Kubernetes CP. Such a service exposes a specific port in every node that runs within the cluster, available to receive requests from outside the cluster.
If pods are replicated, a service performs load balancing among worker nodes that run the replicas. Two types of Kubernetes Services allow external user access: \textit{LoadBalancer} and \textit{NodePort}. \textit{LoadBalancer} requires deployment and configuration of external load-balancing middleware between clients and worker nodes; \textit{NodePort} assigns Kubernetes nodes the load-balancing task. %Both approaches add extra hops on the path between clients and the cluster. With that, \textit{NodePort} combines the control and data paths of cluster users' network flows.
Both \textit{LoadBalancer} and \textit{NodePort} add extra hops on the path between clients and the cluster. While an external load balancer may add operational overheads to manage a cluster, \textit{NodePort} combines the control and data paths of user flows, limiting the scalability of a cluster. These challenges create a necessity for a simpler and more efficient solution to facilitate user access to Kubernetes clusters. 

In this work, we present \textit{P4Kube}, a framework that performs in-network load balancing of a Kubernetes cluster, simplifies a cluster's operation, and improves its performance. P4Kube is implemented on top of Kubernetes \textit{NodePort} Service. Unlike pure \textit{NodePort}, P4Kube decouples the data and the control plane of a Kubernetes cluster, allowing user packets to bypass slower middleware or a Kubernetes CP. P4Kube eliminates the need for an external load balancer, completely offloading this task to the data plane of a router that connects the external network to the worker nodes, and the CP of a Kubernetes cluster. 

%P4Kube updates its state, such as the number and the location of active pod replicas, with the help of Kubernetes CP, which monitors the cluster's current status. For each client packet destined to the IP address and TCP port of a running cluster application, the P4Kube router performs consistent load balancing, overwrites the packet headers, and sends the packet to one of the worker nodes that run the application pod. 

To implement P4Kube, we leverage P4 programming language~\cite{bosshart2014p4}, which allows us to program the data plane of networking devices designed with Protocol-Independent Switch Architecture (PISA)~\cite{sivaraman2015packet}. In P4Kube, this device not only handles load balancing but also performs routing functions, such as Longest-Prefix Matching and packet header overwriting. Additionally, we design a Kubernetes plugin, that monitors a Kubernetes cluster's status and sends control packets to the router data plane. The control packets contain the public port exposed by the \textit{NodePort} Service with which external clients can send requests to reach the internal application, the number of replicas of the application that are running within the cluster, a list of IP addresses of the nodes that these replicas are running on, and the virtual IP address and TCP port for the cluster. P4Kube stores this information in the router's data plane registers. Then, for each client packet destined to the virtual IP address and TCP port of the cluster application, the P4Kube router performs consistent load balancing among the active pods by leveraging the Equal-Cost Multipath (ECMP)~\cite{cai2012rfc} hashing technique. %It overwrites the packet headers and sends the packet to the worker node that runs the selected pods.

We deployed P4Kube using FABRIC testbed~\cite{fabric-2019} and bmv2 programmable switch emulator~\cite{bmv2} and compared it against Kubernetes' original \textit{NodePort} and \textit{LoadBalancer} services. Our results show that for short flows, P4Kube achieves up to 50\% improvement in average request time compared to both services. For longer flows, P4Kube performance is similar to \textit{LoadBalancer} service, and by 20\% faster than \textit{NodePort}.

The contribution of our work is as follows:
\begin{itemize}
    \item P4Kube is a novel approach towards load balancing a Kubernetes cluster with the \textit{NodePort} Service that decouples the control and data paths of a Kubernetes cluster.
    \item P4Kube eliminates the need for an external load balancer;
    \item Our implementation demonstrates the usability of load balancing while leveraging features of a P4 network device, such as custom packet parsing and in-data plane registers.
    \item Our emulations in the FABRIC testbed show up to 50\% improvement of the average request time to the cluster, compared to conventional approaches such as Kubernetes \textit{NodePort} and \textit{LoadBalancer} services.
\end{itemize}

The rest of this paper is organized as follows: In Section~\ref{sec:motivation}, we describe the motivation behind P4Kube in more detail. In Section~\ref{design}, we present the design of P4Kube components, specifically, the Kubernetes CP plugin and the programmable router. Section~\ref{sec:eval} provides the details of the P4Kube evaluation. Section~\ref{sec:related} discusses related work. Finally, Section~\ref{conclusion} concludes the paper.
\section{Motivation}
\label{sec:motivation}
\begin{figure}
	\begin{center}
		\includegraphics[width=0.9\columnwidth]{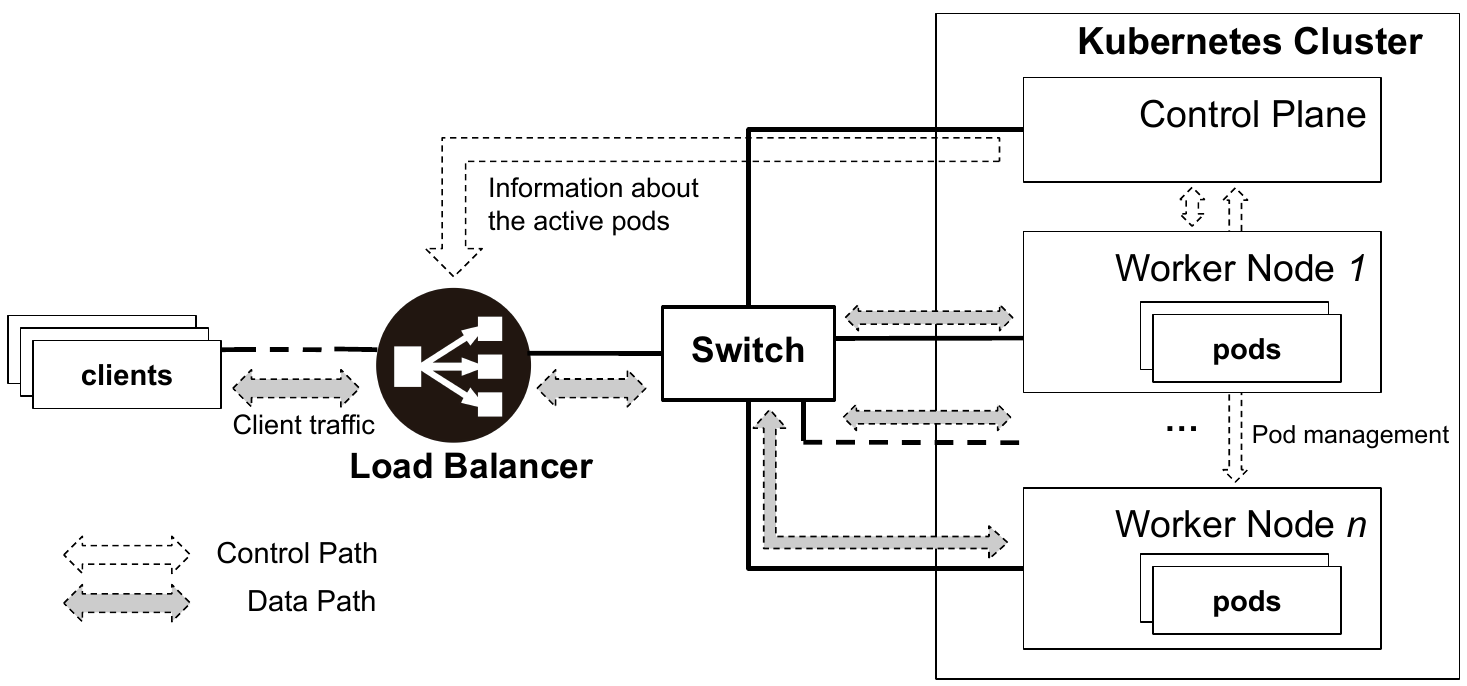}
		\caption{\label{fig:loadbalancer}Kubernetes: \textit{LoadBalancer} Service with an external Load Balancer}
	\end{center}
\vspace{-5mm}
\end{figure}

Within a Kubernetes orchestrator, a Kubernetes Service provides a critical feature of facilitating access to the applications running within a cluster. Among different types of Kubernetes services, \textit{LoadBalancer} and \textit{NodePort} enable external users to interact with a cluster's pods via a public IP address shared among the worker nodes. \textit{LoadBalancer} Service requires the addition of an external load balancer middleware (see Figure~\ref{fig:loadbalancer}). Cooperation between the Kubernetes CP and the external load balancer is required to support correct load balancing when pods are removed or added. Hence, adding an external load balancer requires extra operational and administrative resources. Additionally, a middleware load balancer can create a bottleneck or single point of failure on the data path between the clients and the worker pods~\cite{mesbahi2016load, kogias2020bypassing}.

\begin{figure}
	\begin{center}
		\includegraphics[width=0.9\columnwidth]{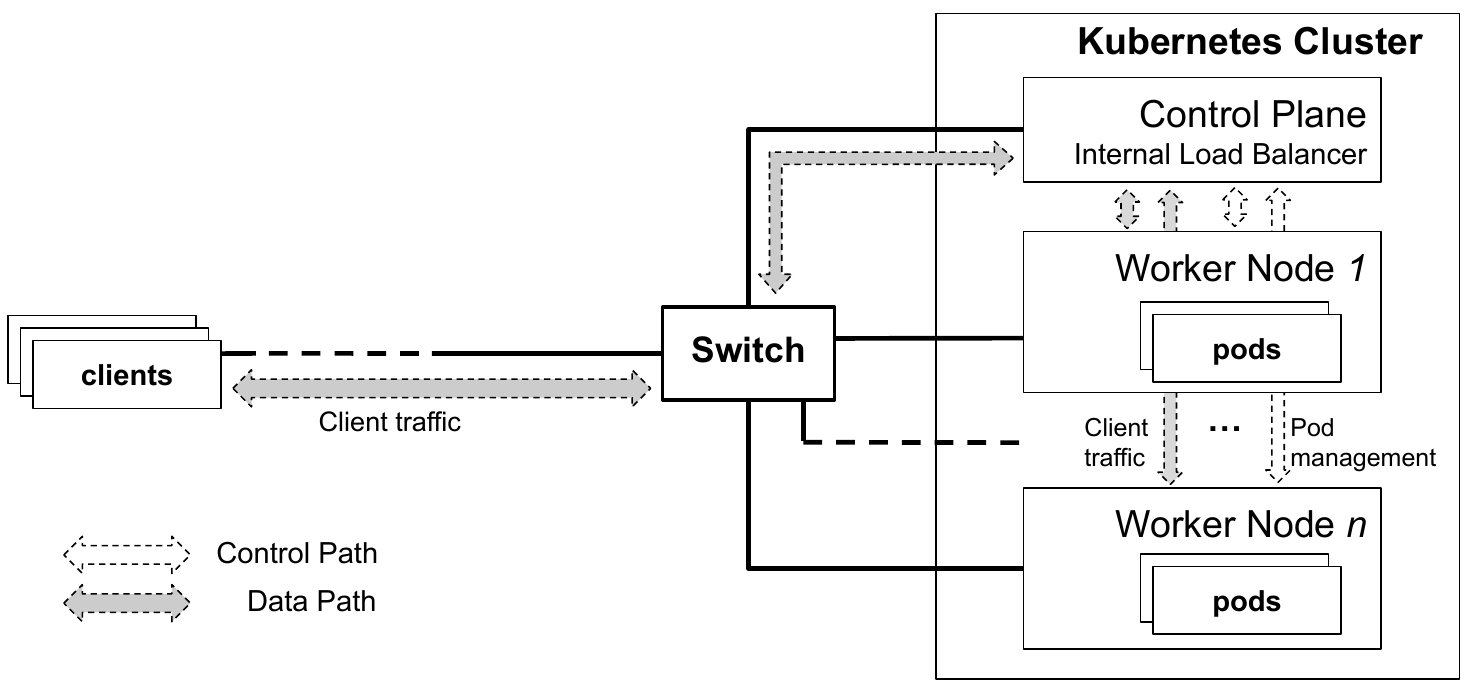}
		\caption{\label{fig:nodeport}Kubernetes:  \textit{NodePort} Service with Kubernetes CP as a load balancer}
	\end{center}
\vspace{-5mm}
\end{figure}

Alternatively, \textit{NodePort} (see Figure~\ref{fig:nodeport}) does not require an extra middleware to expose the cluster pods. To enable load balancing with \textit{NodePort}, the service parameter \textit{externalTrafficPolicy} should be set to \textit{Cluster} (which is the default value set by Kubernetes). In this scenario, the client requests sent to the Kubernetes CP or any other worker node are load balanced by selecting a server pod for each session. The load balancer node re-writes IP headers for ingress and egress packets and routes all packets between the clients and the designated worker nodes. For \textit{NodePort}, Kubernetes uses internal load-balancing algorithms~\cite{ipvs}. While \textit{NodePort} service simplifies the operation of a Kubernetes cluster, it decreases its performance by adding at least one extra hop to the data path between clients and the worker nodes, specifically, the load-balancer node. Even worse, \textit{NodePort} combines data and control paths of Kubernetes, overloading the Kubernetes nodes with the tasks of finding the appropriate worker node and rewriting packet headers. Such a design increases latency and hinders a cluster's robustness in case of Kubernetes' node failures.

In the next section, we describe the design of the P4Kube framework that addresses the aforementioned issues by offloading the Kubernetes load-balancing function to the network data plane located between the clients and the cluster nodes. 
\section{Design}
\label{design}
\begin{figure}
	\begin{center}
 
		\includegraphics[width=0.9\columnwidth]{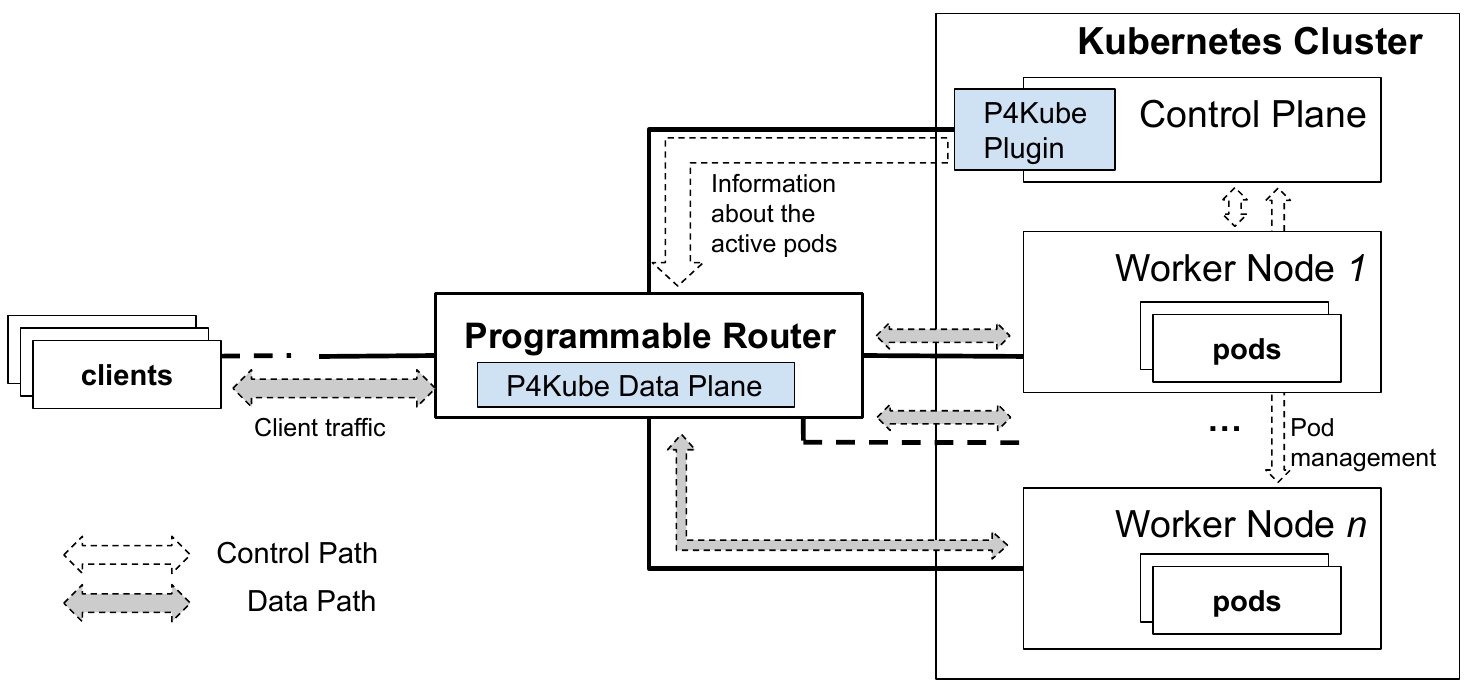}
  %\vspace{-2mm}
		\caption{\label{fig:architecture}System architecture of P4Kube }
 % \vspace{-2mm}

	\end{center}
 \vspace{-5mm}
\end{figure} 
\label{sec:design}
The overall design of P4Kube is demonstrated in Figure~\ref{fig:architecture}. P4Kube consists of two components: (a) the integration plugin, which helps the Kubernetes CP to manage the router, and (b) the programmable data plane program performing load balancing and packet forwarding. Next, in this section, we describe these components in detail.

\subsection{P4Kube integration plugin}
\begin{figure}[h!]
	\begin{center}
		\includegraphics[width=1\columnwidth]{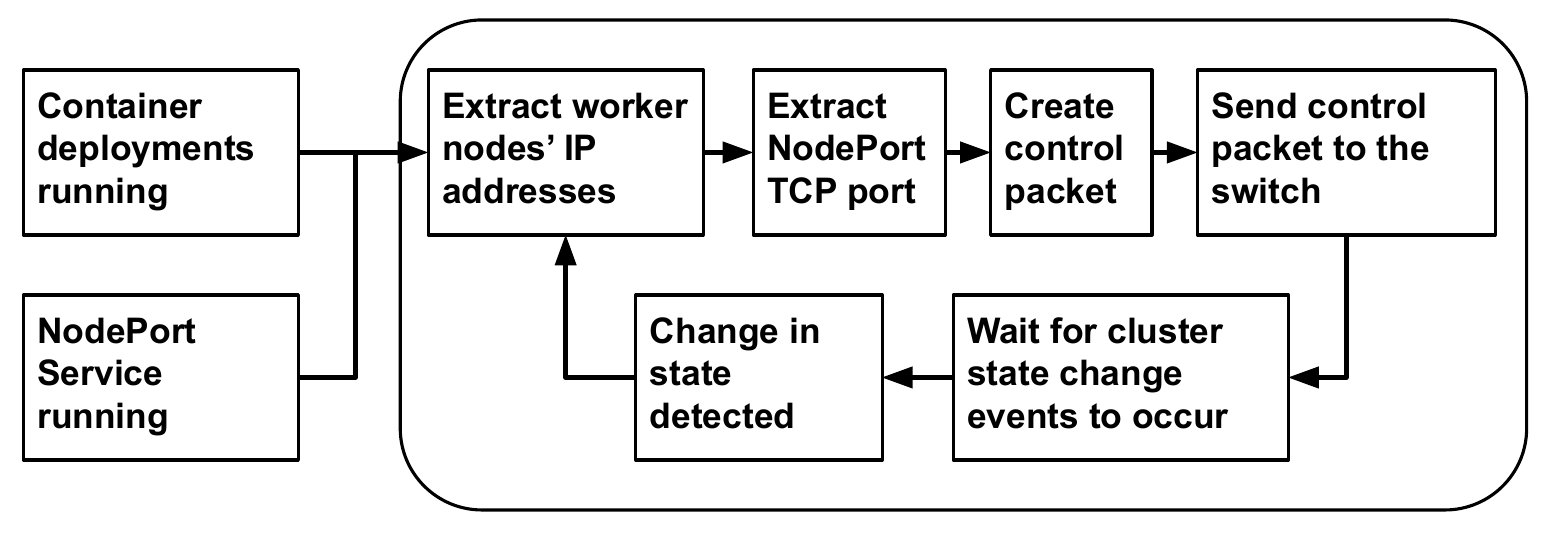}
  \vspace{-5mm}
		\caption{\label{fig:plugin}P4Kube integration plugin workflow}
 % \vspace{-2mm}

	\end{center}
 \vspace{-5mm}
\end{figure}

The novelty of P4Kube is creating an integration between Kubernetes CP and the programmable router to effectively load balance a Kubernetes cluster under changing conditions. To accomplish this, we developed a Kubernetes plugin in the Go language. The plugin's task is to inform the P4 load-balancing program about the current state of the pods running within the cluster. The implementation consists of two main functionalities: detecting when there has been a change in the state of the cluster and sending the control packet to the P4Kube data plane. The overall workflow of the P4Kube plugin is shown in Figure~\ref{fig:plugin}.

\vspace{-1mm}

P4Kube plugin works on top of Kubernetes \textit{NodePort} Service and Calico Container Network Interface (CNI)~\cite{calico}, which allows the Kubernetes CP to communicate with the worker nodes. The plugin can work with other CNI, such as Flannel~\cite{flannel}. Note that when configuring a Kubernetes Service, it is necessary to deactivate the built-in load balancing by setting the parameter \textit{externalTrafficPolicy} to \textit{Local}. Otherwise, the worker nodes receiving requests may forward them to other nodes, adding an additional hop between clients and pods.

\begin{figure}
	\begin{center}
		\includegraphics[width=0.8\columnwidth]{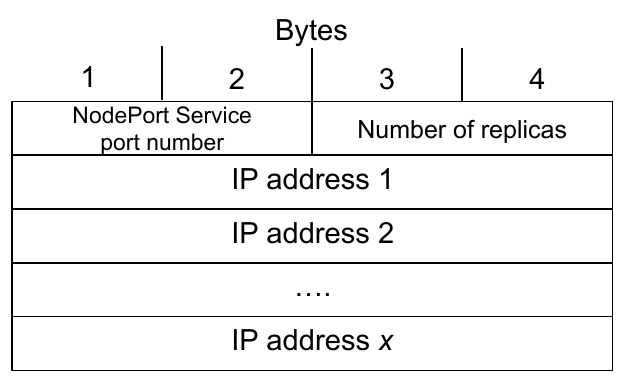}
		\caption{\label{fig:controlpacket}Structure of the P4Kube plugin control packet's payload}
	\end{center}
 \vspace{-5mm}
\end{figure}

On its startup, the P4Kube plugin gathers information about the number of running pods, their IP addresses, and the TCP port exposed by the \textit{NodePort} Service. Next, it forms a control packet and sends it to the programmable router running the P4Kube data plane program. The control packet payload is encapsulated in a UDP header with a specific UDP port (e.g., 7777), allowing the P4Kube data plane to differentiate between data and control packets. See the control packet payload structure in Figure~\ref{fig:controlpacket}.

A containerized system, such as a cluster orchestrated by Kubernetes, is expected to have a dynamic state. Specifically, pods can fail, or new pods can be deployed. The P4Kube plugin detects the changes in the Kubernetes-managed system, forms another control packet, and sends it directly to the P4Kube data plane. Such a design makes P4Kube scalable and robust to failures.

Next, we show how the P4Kube router processes the plugin's control packet, load balances, and forwards data packets between the clients and the Kubernetes cluster pods.

\subsection{P4Kube data plane}
\begin{figure*}
	\begin{center}
		\includegraphics[width=1.7\columnwidth]{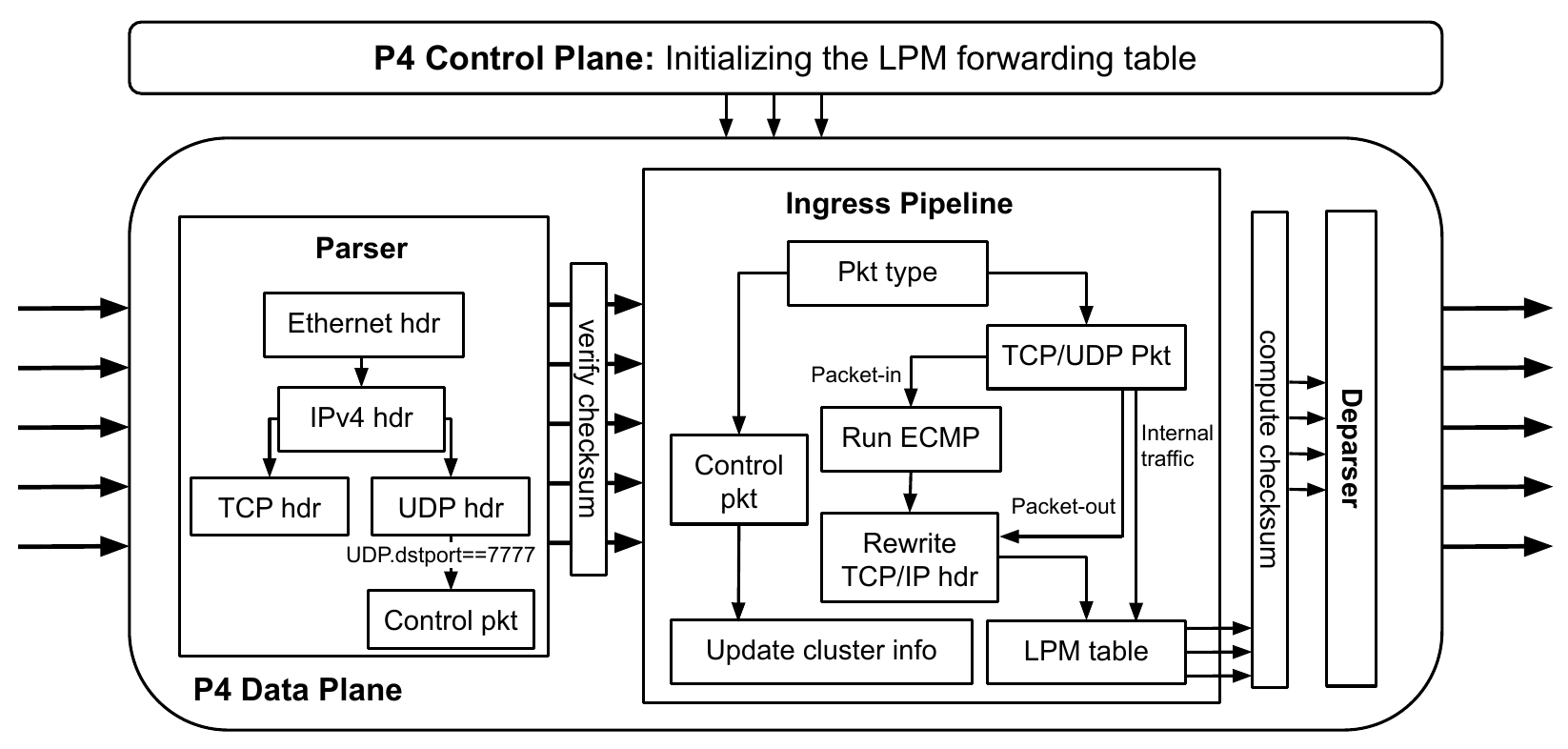}
  %\vspace{-2mm}
		\caption{\label{fig:dataplane}P4Kube data plane workflow }

	\end{center}
 \vspace{-5mm}
\end{figure*} 
To implement the P4Kube data plane program, we leverage P4, a programming language for describing how packets should be processed as they pass through programmable network devices. We use P4 to program the router that connects clients to the cluster. The target of the P4 program is a device designed with Protocol-Independent Switch Architecture (PISA), such as Intel Tofino~\cite{tofino}. The P4Kube router is implemented solely in the data plane; the control plane of the router has a conventional role of updating the Longest-Prefix Matching (i.e., forwarding) table of the data plane, that determines the next hop of a packet based on its destination IP address. 

The P4Kube data plane program is broken down into a series of functions as follows: parser, checksum verifier, ingress processor, egress processor, checksum computation, and de-parser. Figure~\ref{fig:dataplane} shows the workflow of the P4Kube data plane. To demonstrate the ability and efficiency of a P4Kube program while simplifying the problem of load balancing a Kubernetes cluster, we implement load balancing only on data packets that use IPv4 and TCP protocols. Therefore, any mention of ports or port numbers within a packet will refer to the ports of the TCP header unless stated otherwise. In the meantime, P4Kube can be extended to include IPv6 and different transport layer protocols. Next, we describe each component of the P4Kube data plane.

\subsubsection{Parser}
During parsing, packet headers, such as Ethernet, IPv4, TCP, or UDP, are extracted into user-defined data structures. In addition, if the packet is a control packet sent by the P4Kube plugin (see Figure~\ref{fig:controlpacket}), the parser extracts the number of the replicas available in the cluster, their IP addresses, and the port exposed by Kubernetes \textit{NodePort} Service. The control packet is detected via an operator-defined destination UDP port.

After parsing each packet, the data plane verifies its checksum, and if no errors are detected, the packet is passed to the ingress pipeline of the router.

\subsubsection{Ingress Pipeline}
When a packet enters the ingress pipeline, its headers have already been extracted into a usable data structure during parsing. If the packet is a control packet sent from the Kubernetes CP by the P4Kube plugin, the data extracted from the packet's payload is stored in the data plane registers of the router.

The information contained within this payload includes dynamic information pertinent to load balancing the Kubernetes cluster, such as the \textit{NodePort} port number, the total number of replicas, and the list of the IPv4 addresses of the nodes that the replicas are running in. This list of addresses is of length equal to the number of replicas field. If two or more replicas run on the same worker node, the duplicate IP addresses are included to achieve the weighted load balancing when handling the incoming traffic. By loading the control packet payload into persistent memory, this information can continue to be accessed while processing packets in the future. The variable length list of IP addresses is handled by an ``if-else" block of code that considers each possible length up to a defined maximum length, in place of P4 having no support for variable length data structures or ``for" loops. 

For data packets, P4Kube considers three possible cases: 

\begin{itemize}
    \item Incoming packets to the Kubernetes cluster;
    \item Internal packets of the Kubernetes cluster;
    \item Outgoing packets from the Kubernetes cluster.
\end{itemize}

For the first category, P4Kube performs the load-balancing operation to choose the appropriate pod to receive the packet. The incoming traffic is determined by the destination IP address and the destination TCP port of the packets, which should match a cluster virtual IP address and port. 

Once P4Kube determines that a packet falls under incoming traffic, it balances the traffic across multiple nodes that actively run a pod or several pods. P4Kube load balancing leverages Equal-Cost Multi-Path (ECMP) hashing technique. The implementation includes performing a Cyclic Redundancy Check 16 (CRC16) hash of the source IP address, destination IP address, IP transport protocol value, source TCP port, and destination TCP port. The hash value is then used as a lookup value in the register containing the IP addresses of candidate worker nodes. An important aspect of this strategy is that the hash is determined from packet fields that remain consistent for the entire duration of a flow. By hashing these values, we ensure that traffic from the same external source reaches the same destination, maintaining TCP session affinity. 

With the new destination obtained, P4Kube overwrites the packet's destination IP address and TCP port. The virtual IP address is replaced with the address of the selected worker node, and the public port is replaced with the \textit{NodePort} port number. The packet’s next hop is then determined by a lookup on the LPM table using the new destination IP address. 

For the internal packets, i.e., packets that begin and end within the cluster, no load balancing needs to be performed. This traffic only needs to continue to be routed by performing a lookup on the routing table of the device. Internal packets include packets sent by worker nodes to each other or packets to and from the Kubernetes CP.

Finally, if a packet is an outgoing packet sent by a worker node to a client, P4Kube rewrites the packet's source IP address and TCP port values to match the public virtual IP address and TCP port known by the clients. The packet is then matched against the LPM table to determine the next hop. Similarly to the case with the internal packets, no load balancing is performed on such traffic.

\subsubsection{Checksum Computation}
After ingress processing, the checksum values within IP and TCP headers have to be updated to accommodate the changes made to these headers by the P4Kube data plane program. The checksums are recalculated with the built-in functions of a programmable router.
\subsubsection{De-Parser}
During de-parsing, the extracted and updated headers are added back to the packet before the outgoing packet is forwarded.

%With the combination of the P4Kube plugin application sending the control packet and the load balancer parsing and storing the information stored therein, we successfully demonstrate the capabilities of a P4 approach to load balancing a Kubernetes cluster. Unlike the previous designs, the P4Kube packet processing and load balancing eliminates the need for an external load balancer or network control plane.

By combining the P4Kube plugin with the data plane load balancer, we designed a framework that eliminates the need for an external load balancer or network control plane for external traffic forwarding in a Kubernetes cluster. In the next section, we evaluate the performance of P4Kube.
\section{Evaluation}
\label{sec:eval}
\subsection{Experimental setup}

\begin{figure}
	\begin{center}
		\includegraphics[width=0.9\columnwidth]{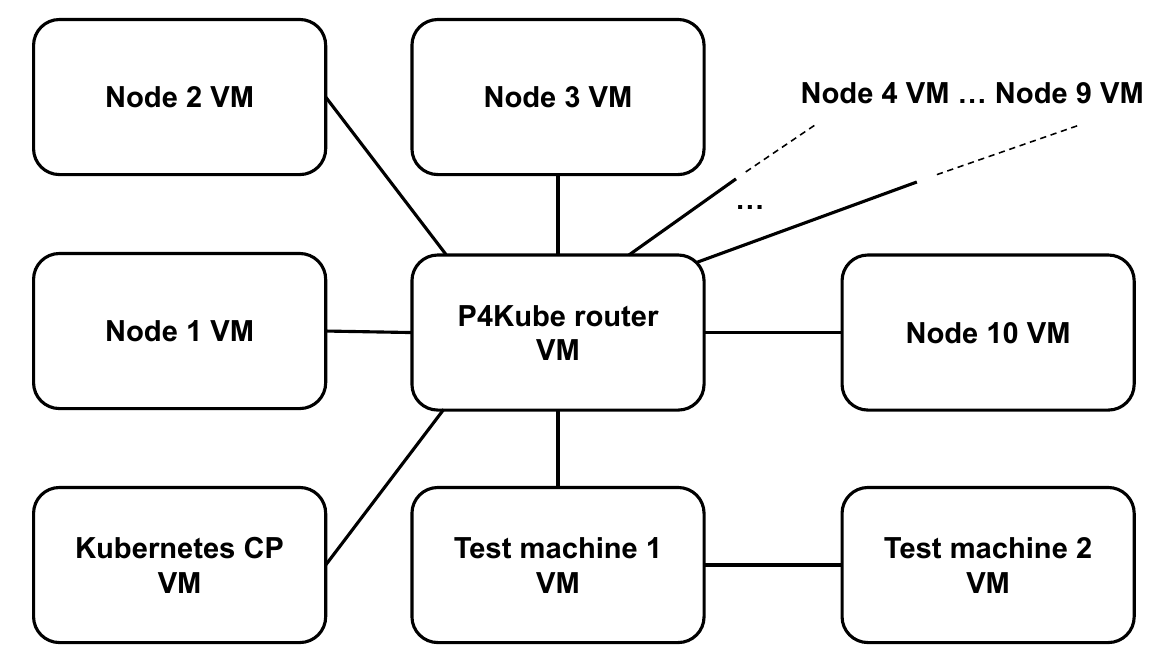}
		\caption{\label{fig:slice}Experimental setup on the FABRIC testbed~\cite{fabric-2019}}
	\end{center}
 \vspace{-5mm}
\end{figure}

\begin{figure*}
	\begin{center}
		\includegraphics[width=2\columnwidth]{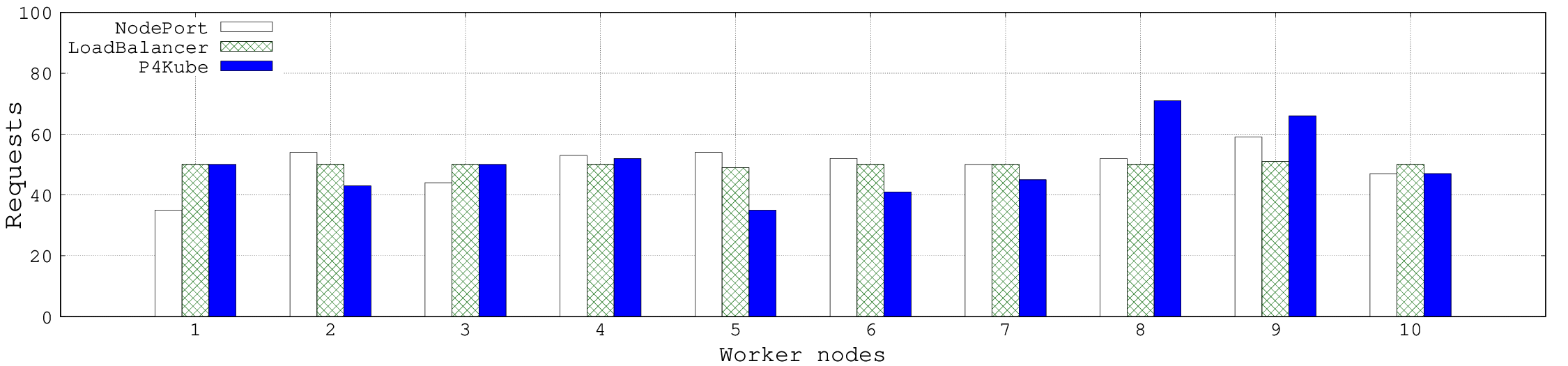}
		\caption{\label{fig:lbresults}Load balancing of 500 concurrent requests}
	\end{center}
 \vspace{-5mm}
\end{figure*} 

\begin{figure}
	\begin{center}
		\includegraphics[width=0.9\columnwidth]{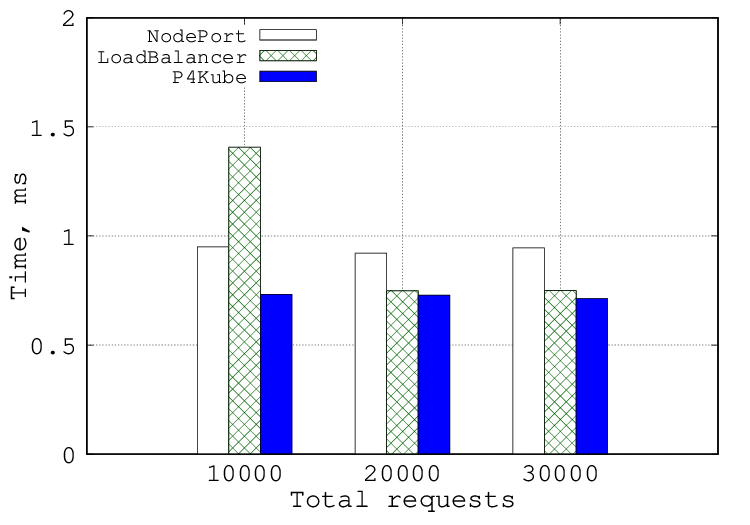}
		\caption{\label{fig:timeresults}Average request time for 500 concurrent clients}
	\end{center}
 \vspace{-5mm}
\end{figure}

We designed P4Kube prototype in P4 and Go programming languages. To evaluate P4Kube, we emulated a realistic cluster using a FABRIC testbed slice~\cite{fabric-2019}. Our experimental setup consisted of VMs comprising a cluster with ten worker nodes, the Kubernetes control plane node, the router, and two additional machines (see Figure~\ref{fig:slice}). All VMs were running Ubuntu 20 and were provided with two CPU cores and 6GB RAM. The router node was running bmv2~\cite{bmv2}, a software emulation of a P4 switch. Note that the bmv2 switch emulator has limited performance and is not meant to measure the performance of the P4 program. Instead, it is used as a proof-of-concept tool for developing and testing programs for hardware switches, such as Tofino~\cite{tofino}, that use high-performance ASICs to run the P4 program for each incoming packet. Nevertheless, in this experiment, we compare the performance of regular Kubernetes services versus P4Kube to estimate possible overheads of the P4Kube plugin, as well as load balancing and packet header overwriting inside the data plane.

For testing purposes, we deployed the NGINX web server~\cite{reese2008nginx} in the Kubernetes cluster, one replica pod per each of the ten worker nodes. To measure the performance of the cluster, we used Apache HTTP server benchmarking tool~\cite{ab} to concurrently send thousands of GET requests, simulating a real-world scenario. Finally, to compare P4Kube against Kubernetes \textit{LoadBalancer} Service with an external load balancer, we used the production-grade NGINX load balancer~\cite{lb} and installed it on the \textit{Test machine 1} to load balance requests from the \textit{Test machine 2}.

\subsection{Load balancing}
First, we compared the load balancing of P4Kube against \textit{NodePort} and \textit{LoadBalancer} Kubernetes Services. We simulated 500 concurrent HTTP requests to the cluster and calculated the number of responses from each worker node. Figure~\ref{fig:lbresults} shows the results of the load-balancing experiment. By default, \textit{LoadBalancer} Service runs the Round Robin algorithm, which results in an equal number of requests shared among the worker nodes. \textit{NodePort} Service load balances the traffic using its Kubernetes internal algorithms such as IPVS.

P4Kube load balances the HTTP traffic using the ECMP hashing. Our results show that ECMP achieves near-optimal load balancing in this scenario. In the meantime, unlike other algorithms like Round Robin, ECMP naturally supports the TCP session affinity requirement. Hence, the P4Kube data plane does not have to store any information about the active TCP connections, making it scalable for a large number of concurrent sessions. %In our earlier work~\cite{grigoryan2023towards}, we implemented a Round Robin-like load balancing in the P4 data plane. Such an approach requires encapsulation of the session ID (or a server ID) into the timestamp field of the TCP option of each packet, except the SYN packet. Other approaches require modification of the Linux kernel~\cite{kogias2020bypassing}.

\subsection{Average request time}

In this experiment, we ran up to 40,000 GET requests through 500 concurrent TCP sessions. We had to limit the concurrency and the number of GET requests to these values due to an increased ratio of non-200 HTTP responses from the \textit{LoadBalancer} Service cluster caused by limited resources of the virtual machines on which we ran the experiments.

Figure~\ref{fig:timeresults} shows the results. It can be seen that P4Kube works best for shorter flows since it selects the designated server using hashing in the data plane, bypassing the control plane, and does not save the state of each session. For 10,000 requests, its average time per request is almost twice as short as when using an external load balancer with Kubernetes \textit{LoadBalancer} Service. Two factors contribute to this difference: (a) Additional hop required for an external load balancer; (b) Initial setup done by NGINX load balancer, required to support the TCP affinity requirement for further packets of each session. With more packets, the difference between P4Kube and Kubernetes \textit{LoadBalancer} diminishes.

Compared to \textit{NodePort} Service, P4Kube works faster by more than 20\% regardless of the flow size. Unlike P4Kube, \textit{NodePort} Service adds two additional hops on the path between the clients and the worker node serving the request, specifically, the Kubernetes CP and the router. 

\subsection{Cluster state change response}
In this set of experiments, we estimated how fast the P4Kube data plane reacts to the changes happening in the cluster, such as a planned de-scaling or up-scaling of a cluster or pod failures. First, we changed the number of replicas via the Kubernetes CP. Within 3 to 5 seconds after the command was issued, the P4Kube plugin sent the control packet to the data plane. Almost simultaneously, the P4Kube data plane updated the number of active replicas and their IP addresses in its registers. Next, we simulated a node failure by turning down the network interface on one of the worker nodes. By default, Kubernetes CP does not immediately update the status of pods on a failed node~\cite{timeout}. In our experiment, it took about 5 minutes for the Kubernetes CP to turn the status of the pod on a failed node into the ``Terminating" state. Once that happened, and an alternative pod was created on a different node, it took 2 seconds for the P4Kube plugin to issue the control packet to the data plane. As in the previous experiment,  once the control packet was received, the data plane immediately updated its registers.

In summary, our experiments show that P4Kube can effectively and efficiently load balance the client traffic to a dynamic Kubernetes cluster. We expect these results to significantly improve when using a hardware switch with a programmable ASIC instead of the bmv2 emulator. In the meantime, the P4Kube prototype currently supports one active external Kubernetes Service with up to ten pod replicas. We plan to extend our prototype in the future, and support other load-balancing techniques, such as load-aware load balancing.

\section{Related work}
\label{sec:related}
In-network load balancing of a Kubernetes cluster has not been thoroughly explored in the existing research. In~\cite{vasireddy2023efficient}, authors present a survey on existing alternative load-balancing solutions for Kubernetes and provide only one reference to a P4-related work. Specifically, Jain et al. in~\cite{jain2021kubernetes} leverage P4 to improve Kubernetes \textit{ClusterIP} Service used for internal communication between worker nodes. They offload internal load balancing and other network services from Kubernetes software to a P4-programmed Infrastructure Processing Unit ASICs, designed by Intel~\cite{ipuintel}. In contrast, P4Kube is designed to offload load balancing among worker nodes for external users of a cluster by implementing it in the programmable network data plane. Authors in~\cite{kogias2020bypassing} design CRAB, a network layer load balancer implemented in P4. CRAB requires modifications of the Linux kernel to bypass TCP affinity requirements for clients. Unlike P4Kube, CRAB does not support dynamic load balancing adjustment when the state of a Kubernetes cluster changes. Works such as~\cite{rizzi2021charon, pit2018stateless, grigoryan2023towards} propose load-aware load balancers in P4 but do not provide integration with a Kubernetes orchestrator. In~\cite{scano2023enabling}, authors use a P4-programmed data plane to collect telemetry for Kubernetes clusters.

\section{Conclusion and Future Work}
\label{conclusion}
%The energy consumption of IT equipment in data centers continues to increase, and mitigating it will benefit the environment and decrease operational expenses. In this work, 
In this work, we propose P4Kube, a system that performs load balancing of external traffic to a Kubernetes cluster. P4Kube bypasses a switch control plane and does not require an external load balancer middleware. We demonstrate that P4Kube improves the average request time to a cluster compared to the conventional Kubernetes services. In addition, it can automatically adjust the load balancing in case pod replicas are removed or added. Future work includes implementing support of multiple Kubernetes services operating simultaneously, and load-aware load balancing via P4Kube.

\bibliographystyle{IEEEtranS}
\bibliography{sample-base}
\end{document}